\documentclass[aps,prl,onecolumn,superscriptaddress,nofootinbib,
nobalancelastpage,nobibnotes]{revtex4}
\usepackage{epsfig}
\begin{document}

\title
{Quantum break in models of axion dark matter }

\author{R. F. Sawyer}
\affiliation{Department of Physics, University of California at
Santa Barbara, Santa Barbara, California 93106}

\begin{abstract}
A system of light axions comprising a classical axion field, one candidate for dark matter, has an instability that can rapidly mix in photon pairs in a coherent fashion if initiated by a quantum break (which eliminates the need for seeds.) Adding more photon states, such as a multiplicity of angles for the case of the axion field at rest, reduces the argument of a logarithmic factor in the mixing time by orders of magnitude. Admitting multiple photon states, all within a window of instability, leads to a synchronization effect that appears to nullify conventional red-shift limitations. Even the fully developed states of the electromagnetic field produced are highly non-classical; they can be looked on as quantum superpositions of different nearly-classical macroscopic systems.
 \end{abstract}
 \maketitle

\subsection{1. Introduction}
Cosmological models in which the dark matter is composed of light axions, in an essentially classical condensed state, have attracted attention recently \cite{witten} - \cite{l}. Here we shall look again at the time evolution due to electromagnetic interactions of a piece of this matter, consisting of $N_a$ axions contained within a periodic box
of volume, $V$, and over a time interval somewhat less than the light travel time over the box. We assume a standard interaction,
$\mathcal{L}_I=g_\gamma a \vec E\cdot \vec B$, where $a$ is the axion field. Axion masses and couplings from literature are in the ranges, $g_\gamma=10^{-21} -10^{-22}  ~{\rm e V}^{-1}$  and
$10^{-22} {\rm eV}< m_a <10^{-4}$eV.

From early in the development of this subject, it has been known  \cite{wil}- \cite{l} that in some regions there is an instability in a mean-field approach that could result in mixing that increases exponentially in time, with growth as $\exp[{r_{g} t}] $, where $ r_g\approx g_\gamma (\rho m_a)^{1/2}$ and $\rho =$[energy density], given some initial mixing. In later and recent literature \cite{l1}-\cite{ll} possible consequences of this instability have been explored. Generally, authors assume an axion mode of high enough occupancy to be considered classical and postulate some small initial mixing with a two-photon mode with equal and opposite momenta, attributed to a quantum fluctuation, then following the whole system as though it were classical. 

In contrast, we begin at $t=0$ with a pure axion state at rest, and 
we use an improved form of mean-field approach that contains the quantum mechanisms necessary to avoid assuming some arbitrary ``vacuum fluctuation" as a seed. In few-mode calculations the method can be tested against numerical solutions of the field theories for cases of initial axion numbers up to the thousands. The time development can than be described as a long gestation time of order $r_g^{-1} \log ( \rho  m_a^{-4})$ during which little happens that is apparent, followed by a sudden near-complete transformation of axions into photons. The argument of the logarithm here is an estimate. 

The above type of behavior can be designated a ``quantum break", a term that has gained currency in describing a genre of actual and conjectured phenomena in several areas: in condensed matter literature
describing, e.g. Bose condensates of atoms \cite{va}-\cite{va3} ; in polarization exchange processes in colliding photon beams  \cite{rfs1}-\cite{rfs4} ; in cosmology 
\cite{cos1}-\cite{cos3}; in flavor physics in dense neutrino clouds \cite{rfsx}-\cite{n12}. In each case the initial state is taken to be stable in a mean field theory (MFT). In each case there is a well defined break-time. For a case with a large number, $N$, of particles it is generally found that the time waiting for the break is of the order of $g^{-1} \log N$, with definitions such that $g$ is a generic coupling strength and cross-sections would be proportional to $g^2$.

\subsection{2. The model.}
We define $c_{j}^\dagger$, $d_{j}^\dagger$ to create  photons with respective momenta ${\vec q_j}$ and $-\vec q_j$ of the same helicity, and $b$ to annihilate an axion in the original condensed mode, where we take the initial occupation number of the mode to be $N_a$ and the axion cloud at rest. The $i$ will be used to index a set of $N_d$ final states with photons that have various magnitudes and directions, but always appear in pairs of opposite momenta.
We write an effective interaction Hamiltonian that describes the mixing induced by the Hamiltonian,
\begin{eqnarray}
&H_{\rm eff}= {\lambda \over ({\rm vol.})^{1/2}}  \sum_{j}^{N_d}   [  b \,c_{j}^\dagger  \,d_{j}^\dagger+
b^\dagger c_{j}  \,d_{j}  ] 
& + \sum_{j}^{N_d}(\omega_j) ( c_{j}^\dagger  \,c_{j}\label{ham}+d_{j}^\dagger  \,d_{ j})\,,
\nonumber\\
\,
\label{hamo}
\end{eqnarray}
where in the last line, the photon kinetic term, $\omega_j=|\vec q_i |-m_a/2$ is the energy of the photon in excess of the conserving value.

We introduce new operators defined in terms of $b,c_j, d_j$ above:
\begin{eqnarray}
Z=b~;~Y_j=c_{j}d_{j}~;~X_j=c^\dagger_{j}c_{j}+d^\dagger_{j}d_{j}\,,
\end{eqnarray}
The Hamiltonian (\ref{hamo}) is then ,
\begin{eqnarray}
H={\lambda \over ({\rm vol.})^{1/2}}\Bigr [\,Z^\dagger\sum_j ^{N_d}Y_j+Z \sum_j^{N_d} Y_j^\dagger\, \Bigr ] \,+\sum_j^{N_d} \omega_j X_j \, .
\label{ham2}
\end{eqnarray}
The non-vanishing commutators of the new variables with each other are, 
\begin{eqnarray}
[Y_j,Y_k^\dagger]=\delta _{j,k} (X_k+1)~~~,~~~[X_j,Y_k]=-\delta_{j,k} 2Y_j~~~,~~~[Z,Z^\dagger]=1\,.
\label{newcoms}
\end{eqnarray}
The commutators (\ref{newcoms}) along with (\ref{ham2}) give the Heisenberg equations, at which point we rescale according to:
 $x_j=X N_a^{-1}$ , $y=Y_j N_a^{-1}$, $z=Z N_a^{-1/2}$, $s=t \lambda V^{-1/2} N_a^{1/2}=t \lambda n_a^{1/2}$, obtaining,
\begin{eqnarray}
i{d\over ds }z\, = \sum_j^{N_d}y_j\,,
\nonumber\\
i{ d\over ds }y_j= (N_a^{-1}+x_j)\, z\,+2 \bar \omega_j y_j \, ,
\nonumber\\
i{d\over ds } x_j=2( z y_j^\dagger - y_j z^\dagger \,).
\label{mmfes}
\end{eqnarray}
where $n_a$ is the number density of axions and $\bar \omega_j=\omega_j \lambda^{-1} n_a^{-1/2}$ is the scaled energy  \underline{difference} between a photon of energy $m_a/2$ and the energy of the photon beam indexed with $j$ in a multi-beam calculation. 

We revert temporarily to the single mode case $N_d=1$
and find the region of $\bar\omega_1$ in which we have exponentially growing modes. We linearize the ${d x \over ds}$ and 
${dy\over ds}$ equations by setting $z=1$, ignoring the $N_a^{-1}$ term in (\ref{mmfes});  then adding an explicit $ {d y^*\over ds}$ equation.  Solving for the eigenvalues $\lambda$ of the 3$\times$3 matrix of coefficients, we obtain 
$\lambda=0 ,\pm \sqrt {4-\bar\omega^2}$. There are growing modes when $|\bar\omega | <\sqrt 2 $.

If we had followed a conventional, mean-field (or classical) approach starting from the Heisenberg equations for the field operators $ \{b(t), c_j(t) ,d_j(t) \}$, instead of $X_j,Y_j,Z_j,$ the operator equations would be the equivalent of (\ref{mmfes} )if we ignored the $N_a^{-1}$ in the latter group (which has an implicit extra factor of $\hbar$).  But when we assume that we can replace a product of dependent variables on the RHS by the product of their expectation values, we then have two different approximations, depending upon which set of variables we have chosen. They can lead to totally different conclusions. Later we summarize the (to us) overwhelming case for accepting the above choice, which we denote the ``modified mean-field approach" (MMF), in distinction to the ``mean-field" approach based on  $ \{b(t), c_j(t) ,d_j(t) \}$, that is the basis of previous literature. The results will be different in important ways, despite the Lyapunov exponents, which govern the early development, being the same.
 
Now for the case $N_d=1$ we compare the solutions of (\ref{mmfes}) to results of a calculation of the complete wave function $\Psi$ in a system with $N_a$ particles in a 1D box.  The basis of the space is now labeled by a single index $\alpha$ that denotes the number of axions remaining (since right-moving and left-moving photon occupancy is determined from this single number), where $\alpha=0,1.....N_a$ . The non-vanishing matrix elements of the Hamiltonian operator in this space are

\begin{eqnarray}
\langle \alpha+1| H | \alpha  \rangle ={\lambda \over ({\rm vol.})^{1/2}} \alpha (N_a-\alpha +1)^{1/2}\,.
\end{eqnarray}

 The initial wave function is now $\Psi_\alpha=\delta_{ \alpha, N_a} $ and we directly solve the 3N$_a$+3 nonlinear ODE's for a set of values from $N_a=128$, going by factors of two to $N_a=1024$.  In fig. 1 the solid curves show the persistence probability, $\zeta(t)$ for an  axion in the original cloud, against the scaled time variable  $s=t \lambda V^{-1/2} N_a^{1/2}=t \lambda n_a^{1/2}$, where $n_a$ is the number density. These are the solid curves in fig. 1. We note the equal spacings, which indicate a 
 $\log [N_a]$ factor in the turnover times.

 Turning to the MMF approximation to the system, for these small values of $N_a$,
the initial conditions for our problem are now $\langle z\rangle=1$; 
$\langle x \rangle=\langle y \rangle=0$.  In the dashed curves of fig. 1 we show the time dependence of the residual axion fraction 
$\zeta(t)=N_a^{-1}\langle b^\dagger b\rangle$  derived from solutions of (\ref{mmfes}) using the above initial condition, for the same sequence of values of $N_a$ as used above. 

\begin{figure}[h] 
 \centering
\includegraphics[width=2.5 in]{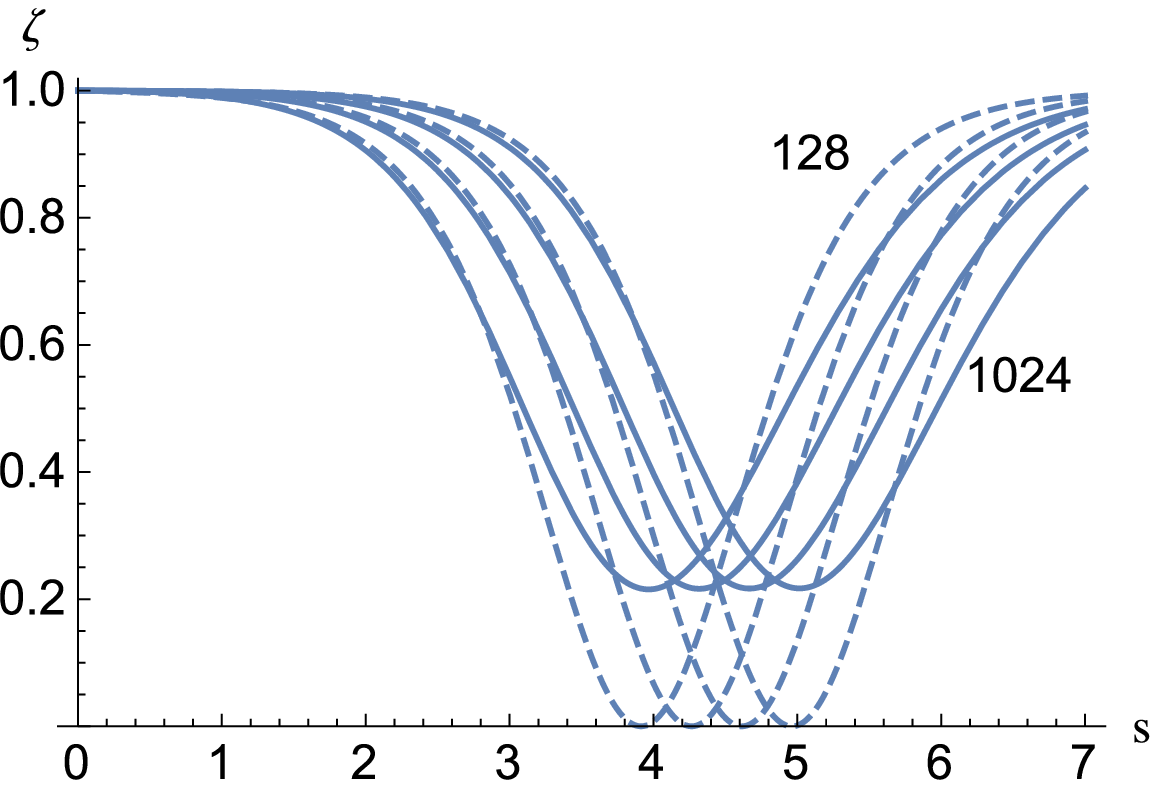}
 \caption{ \small }
Evolution in scaled time $s$. $\zeta$ is the persistence probability for an axion. The solid curves are the result of the solution of the Schrodinger equation for $N_a=128, 256, 512,1024$ going from left to right. The dashed curves are the MMF solutions for the same values of $N_a$.
\end{figure}
Noting the excellent zero-parameter fits shown here from turn-on to the region of large mixing, more or less at
 $ T\approx (\lambda n_a^{1/2})^{-1}  \log_{10} [N_a]$, and for all the values of $N_a$, we seek to compare with the ordinary mean-field approach. But in the latter case we must specify an initial mixing parameter. Experimenting with such choices does not improve on the above zero parameter fit.  

With credibility for the MMF coming from the above comparisons, we can now confirm the logarithmic factor (or nearly logarithmic factor) at much higher values of $N_a$ in the MMF equations, as plotted in fig. 2.
\begin{figure}[h] 
\includegraphics[width=2.5 in]{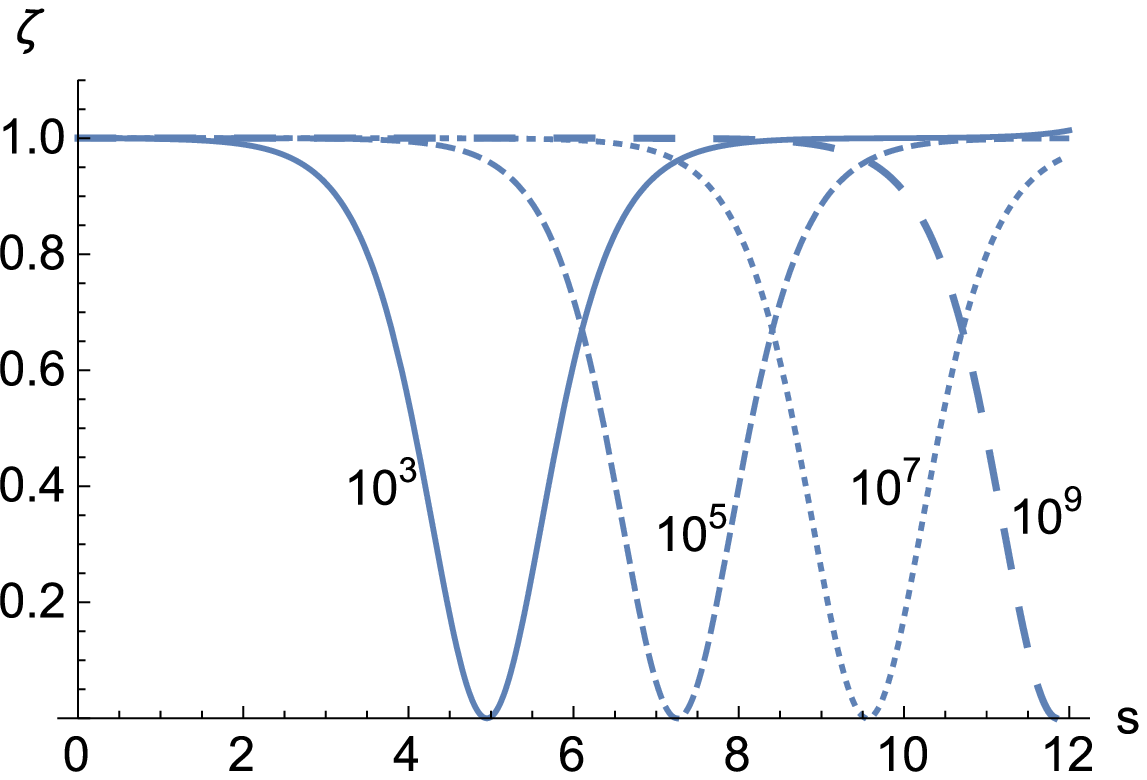}
 \caption{ \small }  Modified mean-field behavior, as in the dotted curves in fig1, but for values of $N_a=10^3, 10^5,10^7, 10^9$.
 \label{ fig. 2}
\end{figure}
 
 \subsection{3. Multiple mode solutions}
First we use the multiple beams ($j=1...N_d$) model solutions of (\ref{mmfes}) and conserve energy exactly, $\bar\omega_i=0$, as would be the case if the modes were to be a spherically symmetrical distribution of $N_d$ rays. This leads to a replacement of the $\log N_a$ factor in the turnover time by $\log [N_a/N_d]$, with the other factors essentially unchanged, as long as $N_d<<N_a$. We point out that this is a very different solution to the problem of what happens when an axion cloud at rest decays in 3D than do those proposed in \cite{hz}, which end up with complex angular distributions. The latter approach is constrained by its formalism to demand that the resulting configuration be a classical E\&M field. Our MMF approach, while not completely satisfactory, accommodates a quantum superposition of rays in the full $4 \pi $ solid angle. Full isotropy of the state is then no problem, even in the presence of translational invariance.

More generally, estimating $N_d$ also involves the counting of all modes that meet the periodic boundary conditions on the surface of the big box and conserve energy to within $\Delta E ~T<<1$ where $T$ is the duration of our calculation. The estimation involves both a multiplicity of photon directions and small deviations in transverse absolute momenta. We have estimated the net result as being the replacment of the above logarithm by  $\log ( \rho  m_a^{-4})$ where $\rho$ is the energy density of axions.

We can go farther and introduce an energy spectrum through the $\bar \omega_j =(E_j^{(\gamma)}-m_a/2)\lambda n_a$ term in the $\dot y_j$ equation from (\ref{mmfes}). As an example we solve for a case $N_a=200,000$, and $N_d= 100$ modes with energy range exactly covering the whole unstable region. We plot the results for four modes with equally spaced energies
that span the unstable region in  $\bar \omega=\sqrt 2 \times \{0, .33, .66, 1 \}$, plot the retention $\zeta(t)$, and then compare to the corresponding calculation of  
$\zeta(t)$, using the conventional mean fields $b,c,d$ where the initial mixing in the latter case was chosen to agree with the short-time MMF results in the former case.  Comparison of the plots shows conclusively that the results 
of an MMF calculation can be very different from those of the mean-field, or basically classical, method.
 \begin{figure}
 \centering
\includegraphics[width=2.5 in]{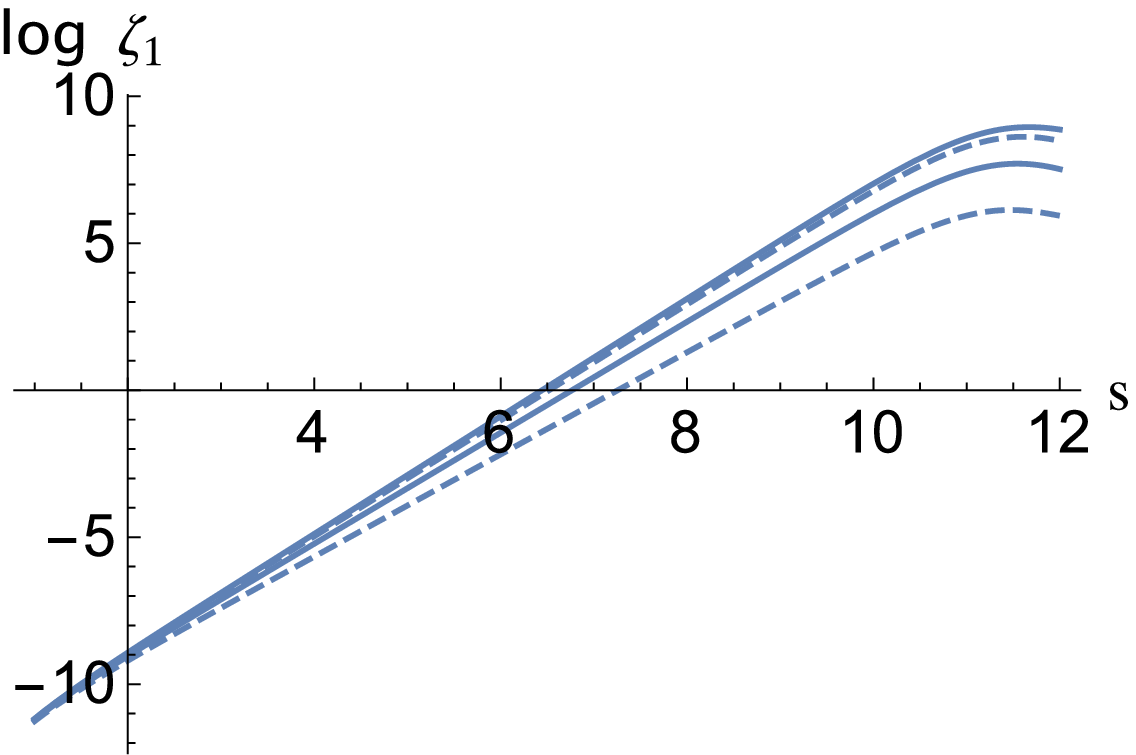}
\caption{ \small } Plot of mode growth in MMF model based on (\ref{mmfes}), where the topmost curve (solid) is for the ``resonant" photon energy $m_a/2$. The three curves counting downwards are for photon energies spanning the complete, unstable region :.
\label{ fig. 3}
\end{figure}

\begin{figure}
\centering
  \includegraphics[width=2.5 in]{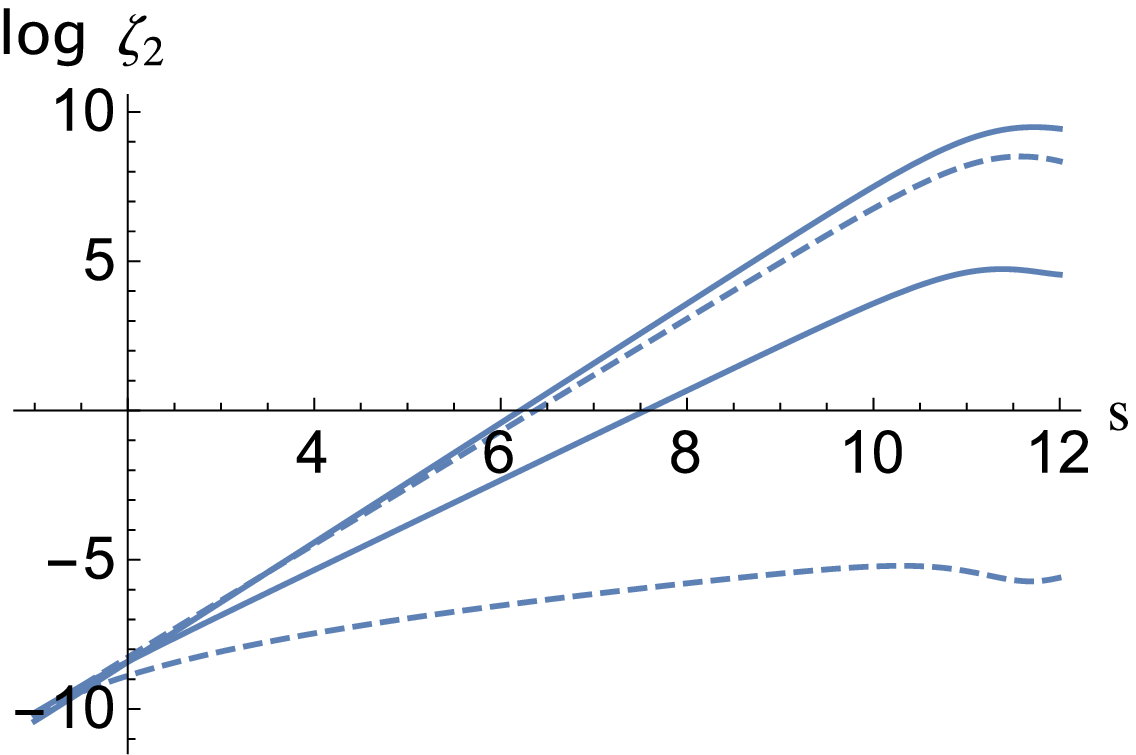}
  \caption{\small} The same as fig. 3, but for the ordinary mean-field, or classical, model. $\zeta_2$ is the persistence probability for an axion.
  \label{fig. 4}
\end{figure}

\subsection{4. Red shift effects}

In a cosmological application we envision ourselves awaiting a pulse of photons that was initiated at a time T in the past, where T is the axion-photon mixing time. We begin by turning on the coupling, taking local t=0, everywhere, at a particular value of the local red-shift. Now the problem is that the initial mixings, being partly at a distance cT to the left of the observer and partly at cT to the right of the observer, for large T can be red-shifted out of the local region of instability long before their rendezvous at the observer's position. The photons become too red and therefore cannot stimulate more extraction of photons from the axion substrate to build the coherent photon field. To address the issue analytically we first define $\xi$ as the fraction of energy that a photon of initial energy $m_a/2$  has lost (as perceived to an observer anchored to the axions) between $s=0$ and $s=S$, where $S=T \lambda n_a^{1/2}$ the scaled mixing time. We must bear in mind that we have to calculate $S$ first (which we estimate in the absence of red shift). We then define 
\begin{eqnarray}
\bar\omega_j(s)=\bar \omega_j(0)(s/S-1/2)\xi
\label{omj}
\end{eqnarray}
which encodes the mismatch between the axion environment and the red-shifted photons in a way that it vanishes midcourse. 
We then write the $ dy/ds$ equation in (\ref{mmfes}) with a term that incorporates this mismatch $\bar \omega_j(s)$, recalling that $\bar \omega_j(s)$ was defined as the scaled difference between the energy of photon $\j$ and the conserving value $m_a/2$. With the explicit time ( $s$) dependence  the evolution equations now read,
\begin{eqnarray}
i{d\over ds }z\, = \sum_j^{N_d}y_j\,,
\nonumber\\
i{ d\over ds }y_j= (N_a^{-1}+x_j)\, z\,+\bar \omega_j(s)\, y_j \,,
\nonumber\\
i{d\over ds } x_j=2( z y_j^\dagger - y_j z^\dagger \,).
\label{step}
\end{eqnarray}

We begin solving (\ref{step}) by looking at the single mode case $N_d=1$. At short times after turn-on any mode within the instability window 
$-\sqrt 2<\bar \omega _j <\sqrt2$ grows independently for a while. But as the red-shift parameter $\xi$ is increased beyond unity, the final mixing at scaled time $S$ quickly becomes negligible, exactly as predicted by the qualitative red-shift argument.

But now we use a large number $N_d$ of modes in the same equations, distributed uniformly throughout the window defined above. As an example we choose a red shift parameter $\xi=5$ that by the criteria above is five times bigger than that which chokes off growth for the single mode, $N_d=1$ case. Choosing $N_a=500,000$ , $N_d=5, 50, 500, 5000$ we plot in fig. 4 the logarithm of the net probability,
$\eta$, for an axion to have turned into photons.
 We define $\eta$ as the logarithm of the net transition probability for an axion to have been converted into photons. The value $\eta=0$ corresponds to total conversion of axions to photons, and it appears that we can come close to this result.
 \begin{figure}[h]
 \centering
\includegraphics[width=2.5 in]{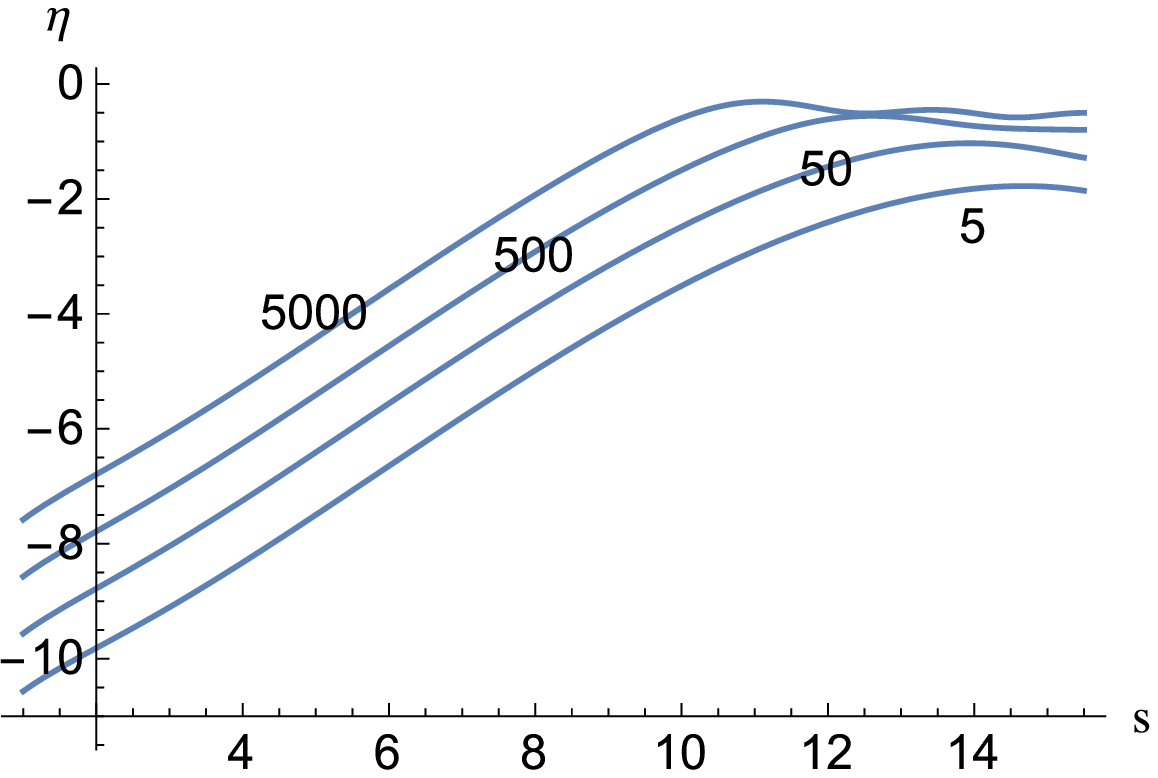}
\caption{ \small } Plot of $\eta=\log_{10} v(s)$ where $v(s)$ is the transition probability for an axion to have been converted to two photons, and $s$ is the scaled time. Curves are labeled by the number $N_d$ of channels assumed in the unstable region of energy.  The number of initial axions is taken as $N_a=500,000$, comfortably greater than $N_d$, which ranges up to 5000,
\label{ fig. 5}
\end{figure}

 \begin{figure}[h]ve
 \centering
\includegraphics[width=2.5 in]{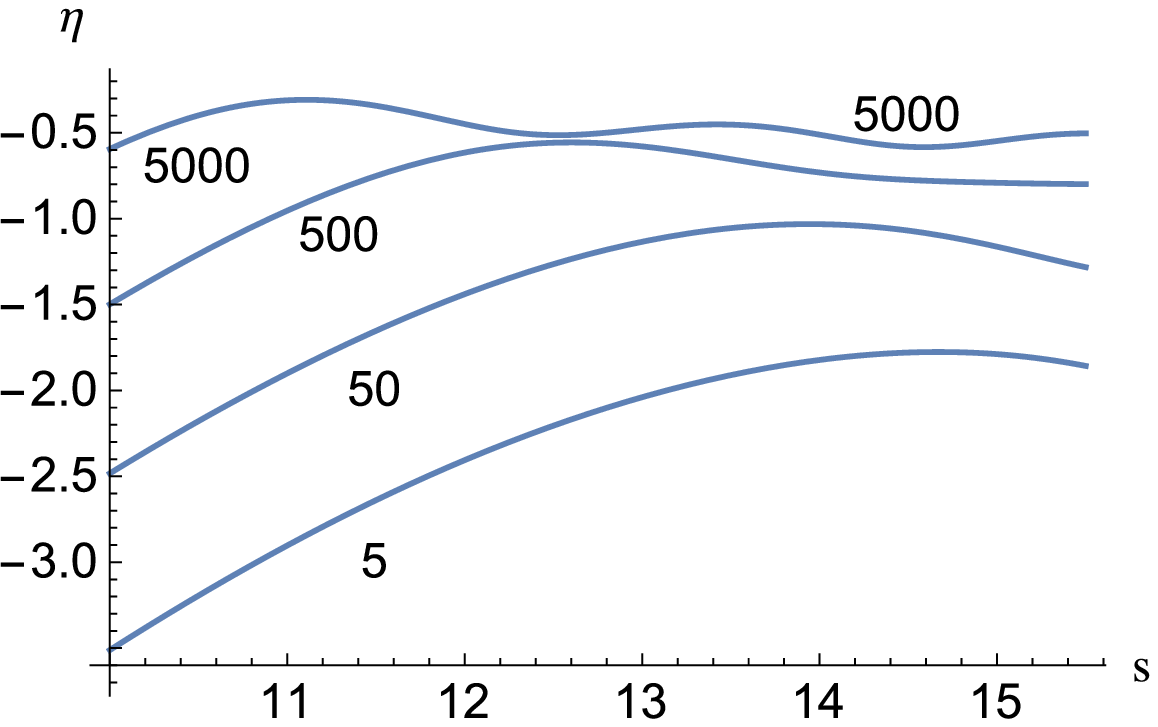}
\caption{ \small }A zoom in on the late behaviors in fig. 5.  Note that the maximum of the $N_d=500$ curve is at a point of near tangency with the $N_d=5000$ curve
\label{ fig. 6}
\end{figure}
Fig. 6 shows a magnified view of these curves in the region of most interest. At maximum, the case with $N_d=5000$ has transformed more than 1/2 of the axions to photon pairs while for $N_d=5$ it is about 1/100 at maximum.This is entirely a result of the red-shift induced time dependence of (\ref{omj}). When we set $t=0$ in
and solve (\ref{step}) all the curves achieve complete transformation at maximum. The positions of the maxima are hardly affected at all however, as one expects from the results of sec. 3.

On the basis of the preceding section of this paper, we might have expected that having $N_d$  final channels for which the energy costs were not exorbitant, instead of just one, would serve only to transform the single channel result by replacing a $\log N_a$ in the time scale into say, $\log[ N_a/ N_d]$. But the salient feature that we see in fig.5 is that as $N_d$ becomes large, and at the same time we take the redshifts into account through the time dependence of (\ref{step}) , a peak of gamma ray conversion emerges at a  time close to that which would have been predicted in the absence of a red shift. 
It is in some language a result of intermediate processes $\gamma_q+\gamma_{-q}\rightarrow {\rm axion }\rightarrow \gamma_p+\gamma_{-p}$, and by virtue of countless small denominators, which get smaller and smaller as we increase $N_d$. 

In any case it is the explicit time dependence of the red-shifting term in (\ref{step}) that causes cloud photons to continuously be promoted into the regions most closely tuned to the axion substrate (within the workings of our very simple $3 \times N_d$  equations). In the end we must count the number of unstable modes that should be included in the calculation; and that fit in our periodic box, much as we described in the last section for the argument of the logarithm there. In any case there will be an ample supply of very nearly energy conserving states coming from pairs with tiny transverse momenta. 

There are multitudes of other examples of synchronization in many-body systems in physics. We have not succeeded in connecting the detailed mechanics of the present case to that of any other, and better understood, system. But a similar statement can be made with respect to a number of other synchronizations as well.  Because in the present example the actual numbers $N_a$, $N_d$ of interest for the application are so much greater than in our simulations, and furthermore that we have treated each elaboration of the simplest $N_d=1$ case by itself--never consolidating into one big equation that includes all of our considerations--everything remains a bit tentative. If we had better computing power we could get an extension into a range of higher numbers that would be helpful in building a stronger case, or in modifying the conclusions.
\subsection{5. Time boxes and inhomogeneities.}  
Next we do a calculation in time boxes, dividing the total time interval $T$ into 20 equal segments, taking $T$ to be of the order of the expected mixing time in the monolithic calculation. We then solve for the complete time region by computing values on each segment boundary to use as an initial value on the following boundary. Thus in the first time interval ($\beta=1$) the produced photons that we eventually will see here at the origin are: produced a distance $cT$ to the left of us and moving to their right; and produced  $cT$ to the right of us and moving to their left. In the next time interval, a little closer on both sides, the two groups of live photons encounter new groups moving in the respective counter-directions, but these new counter-groups have been seasoned exactly as have the groups that escaped, and the little piece destined for us doesn't know the difference; and so on, until both initial groups arrive at our location ($\beta=20$) simultaneously, and thoroughly fattened from their interactions with the groups encountered along the way. Up to this point this discrete approach says in the end: ``Save the trouble and take the continuum limit to get back essentially to (\ref{mmfes})."
Now we follow the same mechanics, but let the axion substrate's effective coupling to the axions vary randomly by 20\%, box to box, as our two test segments successively move towards us. These variations are formulated in time boxes, but can just as well be viewed as characteristic of the space regions our respective test segments occupied at a particular time. The result of the calculations is that there is very little effect on the mixing times in most throws of the dice. 

 \subsection{6.Discussion }
 
In our opinion the seeded, and otherwise totally classical, approach has fatal drawbacks. First: the state of the photon cloud is such that each $ \vec q $ photon has as its own mate a $- \vec q$ photon with the same phase. Indeed, as soon as we produce only a few photon pair states from the photon vacuum into a particular mode we can ask the quantum calculation: ``what is the expectation of the electric field in some small region?'', obtaining the answer ``zero", since $\langle c(t)\rangle=0$, $\langle d(t)\rangle=0$, at all times when we start from the pure coherent axion state.  But if we ask instead ``what is the expectation of the photon energy density" we get back the right number for the number of photons. These results cannot coexist for classical fields, but do so trivially for quantum states. 
 
Now consider the difference between the use of our $H$ as an operator, alternatively as a function of the ordinary operators $ b^\dagger, c^\dagger, 
d^\dagger$ and their adjoints, acting in their usual space of states; or as a function of our composite operators $x, y, y^\dagger, z, z^\dagger$. The total space created by the first set is much bigger than that of the second, but only a tiny fraction of these states even connect dynamically, through $H_{\rm eff}$ to our initial axion state; whereas in the second case all of the operative space is connected to the axion substrate.  This is a more promising beginning.
  
Next consider from this viewpoint a basic conundrum of some of the axion decay literature: how do we achieve isotropy in the decay of a homogeneous isotropic axion cloud at rest,
 in spite of the fact that there exists no translationally invariant and isotropic state of the $\mathcal{E \& M}$ field? For some contortions relating to this question in the MFT see, e.g.,
 ref. \cite{hz}, with its vector spherical harmonics. But in the MMFT the clear answer is that we only have to allow the built-in mechanics of our equations (\ref{mmfes} ), now in the multiple directional mode with $N_d$ directions available, but still with 
 $\omega_i=0$. The equations (\ref{mmfes}) will apportion out the pairs in the various directions, and the resulting state is a quantum superposition of these units. 
 
 Added to the above arguments are the illustrations in fig. 3 and fig.4 showing that outcomes of some calculations can be very different in the quantum approach than in a corresponding basically-classical approach, even when the initial seeds for the latter case are fitted to the early time development of the former case. Finally there is the red-shift material of Sec. 4.

We have not arrived at a firm conclusion as to whether when all of the above considerations are taken account there will be a viable application to an axion configuration that could play a role in dark matter cosmology or with respect to compact objects composed of axions. We mention only one rather specific scenario that we believe could be possible, depending on what turn out to be allowable parameter regions. We consider an axion substrate of number density such as to provide all the dark matter, distributed more or less uniformly at the time of recombination. During the reduction in free electron density that leads to the ``dark era", the plasma frequency can fall below 10$^{-11}$eV (using electron densities from \cite{peebles}).  Growth of coherent E\&M fields through the above mixing mechanisms can then begin independently everywhere at that time for the case of an axion mass greater than $\approx 10^{-11}$ eV. In this case we can apply the pure time dependent formalism that led to our evolution equations in the scaled quantities (\ref{mmfes}), but with the $\dot y$ equation amended to include the red-shift effect as in (\ref{step}). With couplings at presently quoted limits we estimate that we would predict large scale photon production on a time scale of a few tens of years after the turn-on, a time still small compared to the horizon scale at recombination. Being totally at odds with observation, this could lead to new parameter bounds. Of course, effects that we have not included could cut off the transformation with only a relatively small number of photons having been produced, enough to be observationally interesting but not enough to veto the theory.

But the speculation is premature; the present work is intended as a development of techniques that are more generally needed {\it en route} to calculating the practical effects of the $a\gamma\gamma $ coupling in systems of high axion number density. 

It is a pleasure to thank Alessandro Mirizzi for much useful correspondence on the issues discussed here.

\end{document}